\documentclass[letterpaper, 10pt, conference]{ieeeconf}

\usepackage{graphics} 
\usepackage{epsfig} 
\usepackage{amsmath} 
\usepackage{amssymb}  
\usepackage{cite}
\usepackage{bm}
\usepackage{subfigure}
\usepackage{acronym}
\usepackage{paralist}
\usepackage{float}
\usepackage{color}
\usepackage{epstopdf}
\usepackage{multicol}
\usepackage{epstopdf}
\newtheorem{critical point}{Theorem}
\newtheorem{Multi-Robot}[critical point]{Theorem}
\newtheorem{Safety-perturbed}[critical point]{Theorem}
\newtheorem{Safety-stoch}[critical point]{Theorem}
\newtheorem{Convergence}[critical point]{Theorem}
\newtheorem{Safe worst case}{Lemma}

\DeclareMathOperator{\F}{\mathrm F}

\DeclareMathOperator{\R}{\mathbb R}

\DeclareMathOperator{\N}{\textrm N}

\IEEEoverridecommandlockouts
\begin{document}

\title{\LARGE \bf Robust Semi-Cooperative Multi-Agent Coordination in the Presence of Stochastic Disturbances}
\author{Kunal Garg \and Dongkun Han \and Dimitra Panagou 
\thanks{The authors are with the Department of Aerospace Engineering, University of Michigan, Ann Arbor, MI, USA; \texttt{\{kgarg, dongkunh, dpanagou\}@umich.edu}.}
\thanks{The authors would like to acknowledge the support of the Automotive Research Center (ARC) in accordance with Cooperative Agreement W56HZV-14-2-0001 U.S. Army TARDEC in Warren, MI and the support of the NASA Grant NNX16AH81A.}}

\maketitle

\begin{abstract}
This paper presents a robust distributed coordination protocol that achieves generation of collision-free trajectories for multiple unicycle agents in the presence of stochastic uncertainties. We build upon our earlier work on semi-cooperative coordination and we redesign the coordination controllers so that the agents counteract a class of state (wind) disturbances and measurement noise. Safety and convergence is proved analytically, while simulation results demonstrate the efficacy of the proposed solution. 
\end{abstract}

\acrodef{wrt}[w.r.t.]{with respect to}
\acrodef{apf}[APF]{Artificial Potential Fields}
\acrodef{ges}[GES]{Globally Exponentially Stable}
\newcounter{MYtempeqncnt}

\section{Introduction}

Coordination in multi-agent systems has attracted much attention over the last decade with a plethora of theoretical and practical problems that this paper can not cite in their entirety; for recent overviews the reader is referred to  \cite{Ren_Cao_2011,knorn16tcns, wang16neuro, mesbahi10pup}. 

A fundamental problem of interest in the area of distributed coordination and control is the decentralized multi-agent motion planning, which mainly focuses on generating collision-free trajectories for multiple agents (e.g., unmanned vehicles, robots) so that they reach preassigned goal locations under limited sensing, communication, and interaction capabilities. Numerous elegant methodologies on planning the motion for a single agent (robot) have appeared in recent years, with the most popular being (i) sampling-based methods, including probabilistic roadmaps \cite{Kavraki_TRA96}, and rapidly-exploring random trees \cite{lavalle2000rapidly, Karaman_IJRR2010}, (ii) Lyapunov-based methods, including either the definition of closed-form feedback motion plans via potential functions or vector fields, or computation of Lyapunov-based feedback motion plans via sum-of-squares programming \cite{Tedrake_IJRR2010, Tedrake_IJRR2014}, and (iii) graph search and decision-theoretic methods, see also \cite{LaValle, Choset} for a detailed presentation. Although each method has its own merits and caveats, arguably Lyapunov-based methods (often termed reactive) are particularly popular for multi-agent motion planning problems, as they offer scalability with the number of agents, and the merits of Lyapunov-based control design and analysis.  

In addition, robustness against modeling and/or measurement uncertainties is of primary importance for real-world systems and applications. Hence the problems of modeling, quantifying, and treating uncertainty are of particular interest when it comes to multi-agent coordination. A straightforward way to model uncertainty in multi-agent systems is by considering them as \textit{bounded disturbances}. In \cite{trentelman13tac}, a stable uncertainty is assumed to be bounded in $\mathcal{H}_{\infty}$-norm by some \textit{prior} given desired tolerance, and a state space observer along with a robust controller are designed by using the algebraic Riccati equations. In \cite{meng13auto}, an additive $l_{2}$-norm bounded disturbance is considered, and a robust controller is proposed for the distributed cooperative tracking problem in a leader-follower network by using Lyapunov stability theorems. In \cite{hu12scl}, both bounded disturbances and unmodeled dynamics are assumed in the dynamics of agents; an identifier for each agent is designed to estimate the unknown disturbances and unmodeled dynamics. Related work considering bounded deterministic disturbances can be found in the design of finite-time consensus algorithms with mismatched disturbances \cite{wang16acc}, and the rotating consensus control with mixed model uncertainties and external disturbances \cite{li15neurocomputing}.  

Another way of modeling uncertainty is by \textit{Gaussian random processes}. In \cite{cheng14tac}, communication noises are described by a standard Brownian motion, and the mean square consensus in the multi-agent system is achieved by proposing a stochastic approximation-type gain vector, which attenuates the effect of noises. This work is extended to networks with Markovian switching topologies \cite{wang15tac}, and leader-follower networks \cite{cheng16tac} with a similar noise-attenuation controller. In \cite{huang09siam}, it is assumed that the measurements for each agent are disturbed by white noises. Robust consensus can be achieved by applying stochastic Lyapunov analysis. Based on this idea, in \cite{li10tac}, the average-consensus problem of first-order multi-agent systems is considered, and a necessary and sufficient condition is proposed for robust consensus by using probability limit theory. The aforementioned methods are efficient in solving the coordination problems with stochastic uncertainties in measurement or system dynamics; however, safety (i.e., the generation of collision-free trajectories) is not considered. 

In contrast to the aforementioned results, in this paper we consider the problem of generating collision-free trajectories for multiple agents in the presence of uncertainty. We propose a robust, Lyapunov-based coordination protocol that achieves collision-free motion for multiple agents in a distributed fashion, in the presence of state and measurement uncertainties. The method builds upon our earlier work in \cite{Panagou_TAC16}, in which the nominal (uncertainty-free) case was considered. More specifically, we redesign the semi-cooperative coordination protocol in \cite{Panagou_TAC16} so that it accommodates the case of state and measurement uncertainties. Our approach yields a method on the safe and robust motion planning of multiple agents that is based on analytic vector fields, hence offers scalability with the number of agents along with provable guarantees. In summary, the contributions of this paper are: (i) a robust, semi-cooperative coordination protocol that accommodates for a class of stochastic disturbances in the agents' dynamics and measurements, and (ii) the derivation of analytical bounds on the navigation (estimation) and final state errors of the agents in terms of the considered uncertainties.

The paper is organized as follows: Section \ref{Modeling} includes an overview of the modeling of the system under the effect of disturbances. Section \ref{Robust Coordination} presents the robust coordination protocol along with the safety and convergence analysis. Section \ref{Simulations} evaluates the performance of the proposed method via two simulation scenarios. Our conclusions and thoughts on future work are summarized in Section \ref{Conclusions}.

\section{Modeling and Problem Statement}\label{Modeling}

Let us consider $N$ identical agents $i\in\{1,\dots,N\}$, which are assigned to move to goal locations of position coordinates $\bm r_{gi}=\begin{bmatrix}x_{gi}&y_{gi}\end{bmatrix}^T$ relative to some global frame $\mathcal G$, while avoiding collisions. The motion of each agent $i$ is modeled under unicycle kinematics with additive disturbances that stand for state and output uncertainty, for instance due to wind effects and sensor imperfections, respectively, as:
\begin{subequations}
	\label{unicycle}
	\begin{align}
	\bm{\dot q}_i&=\bm f(\bm q_i, \bm u_i)+\bm \Gamma\bm w \Rightarrow \begin{bmatrix}\dot x_i \\ \dot y_i \\ \dot \theta_i \end{bmatrix}=\begin{bmatrix}u_i c\theta_i \\u_i  s\theta_i \\ \omega_i \end{bmatrix}+ \begin{bmatrix}w_{x}\\w_y\\0\end{bmatrix},\\
	\bm y_i &= \bm h_i(\bm q_i)+ \bm v_i,
	\end{align}
\end{subequations}
where $\bm{q}_i=\begin{bmatrix}\bm r_i^T&\theta_i\end{bmatrix}^T$ is the state vector of agent $i$, comprising the position vector $\bm{r}_i=\begin{bmatrix}x_i&y_i\end{bmatrix}^T$ and the orientation $\theta_i$ of the agent \ac{wrt} the global frame $\mathcal G$, $\bm u_i=\begin{bmatrix}u_i&\omega_i\end{bmatrix}^T$ is the control input vector comprising the linear velocity $u_i$ and the angular velocity $\omega_i$ of agent $i$, $\bm f(\cdot, \cdot):\mathbb R^3 \times \mathbb R^2 \rightarrow \mathbb R^3$ is the vector valued function of the agent dynamics, and $c(\cdot)\triangleq\cos(\cdot)$, $s(\cdot)\triangleq\sin(\cdot)$ and 
\begin{align*}
\bm{\Gamma} = \begin{bmatrix}
                        1& 0 \\ 
                        0 & 1 \\ 
                        0 &0 
                        \end{bmatrix}. 
\end{align*}
The random process $\bm w=\begin{bmatrix}w_x&w_y\end{bmatrix}^T$ is assumed to be Gaussian, white, of known mean $\bm{\bar{w}}=\begin{bmatrix}\bar w_x&\bar w_y\end{bmatrix}^T$ and known covariance $\bm P_w\in\mathbb R^{2\times 2}$. To maintain the Lipschitz continuity of the proposed control law, we assume that the mean value of state disturbance is continuous in time and is bounded. Furthermore, $\bm y_i\in\mathbb R^m$ is the output vector comprising the available measurements, $\bm h_i(\cdot):\mathbb R^3\rightarrow \mathbb R^m$ is the output function, and $\bm v_i\in\mathbb R^m$ is the measurement noise modeled as a Gaussian, white process of zero mean $\bar{\bm v}_i=\bm 0$ and known covariance $\bm P_{v_i}\in\mathbb R^{m\times m}$. For simplicity in the sequel we assume that the output function is the identity map so that the measurement model reduces to $\bm y_i=\bm q_i + \bm v_i$, and that the measurements are uncorrelated, so that the covariance matrix of $\bm v_i$ reads $\bm P_{v_i}=\mathrm{diag}(\sigma_{v_{i,1}},\sigma_{v_{i,2}},\sigma_{v_{i,3}})$.

Each agent $i$ is modeled as a closed circular disk of radius $\varrho_i$, and has a circular communication/sensing region $\mathcal C_i$ of radius $R_c$ centered at $\bm r_i=\begin{bmatrix}x_i&y_i\end{bmatrix}^T$, denoted as $\mathcal C_i : \{\bm r\in\R^2 \; | \; \|\bm r_i - \bm r\|\leq R_c\}.$ We denote $\mathcal N_{i}$ the set of neighboring agents $k\in \mathcal C_i$ of agent $i$. We assume that each agent $i$ can measure the position $\bm r_k$, orientation $\theta_k$ and receive the linear velocity $u_k$ of any agent $k$ lying in $\mathcal C_i$. 

In our earlier work \cite{Panagou_TAC16} we considered the nominal case of \eqref{unicycle}, i.e., the case for $\bm w=\bm 0$, $\bm v=\bm 0$, and designed the following semi-cooperative distributed coordination protocol:
\begin{itemize}
	\item[\textbf{Coordination of linear velocities:}] The linear velocity $u_i$ of each agent $i$ is governed by the control law:
	\begin{align}
	\label{velocity protocol}
	u_i &= \left\{
	\begin{array}{rc}
	\max\left\{0,\min\limits_{k\in \mathcal N_{i} | J_k<0} u_{i|k}\right\}, & \hbox{$d_m\leq d_{ik}\leq d_\epsilon$,}\\
	u_{i\epsilon}, & \hbox{$d_\epsilon < d_{ik} < d_c$,}\\
	u_{ic}, & \hbox{$d_c\leq d_{ik}$;} \\
	\end{array}
	\right.
	\end{align}
	where:
	\begin{itemize}
	    \item $d_{ij}$ is the Euclidean distance between agents $i$ and $j$, $d_m\geq 2(2\varrho+\varrho_\epsilon)$ is the minimum allowable pairwise distance, $d_c$ is a positive constant such that $d_c\leq R_c$, and $d_r$ is a positive constant such that $d_m<d_r<d_c$,
		\item $u_{ic}=k_{ui} \tanh(\|\bm r_i - \bm r_{gi}\|)$, $k_{ui}>0$,
		\item $u_{i\epsilon}$ is the value of the linear velocity $u_i$ of the agent $i$ when $d_{ij}=d_c$, that is, $u_{i\epsilon}=u_{ic}|{_{d_{ij}=d_c}}$,
		\item the distance $d_\epsilon$ is set equal to $d_\epsilon=d_r-\epsilon$, 
		\item $u_{i|k}$ is the safe velocity of agent $i$ \ac{wrt} a neighbor agent $k\in\mathcal N_{i}$, given as:
		\begin{align}
		u_{i|k}&=u_{i\epsilon}\;\frac{d_{ik}-d_m}{d_\epsilon-d_m}+\varepsilon_i \; u_{is|k} \; \frac{d_\epsilon-d_{ik}}{d_\epsilon-d_m},
		\end{align}
		with the terms in \eqref{safe velocity} defined as:
		\begin{align*}
		u_{is|k}&= u_k \;\frac{{\bm r_{ki}}^T\bm \eta_k}{{\bm r_{ki}}^T\bm \eta_i},\;\; \bm \eta_i=\left[\begin{matrix}\cos\theta_i\\\sin\theta_i\end{matrix}\right],\;\; J_k={\bm r_{ki}}^T\bm \eta_i, \\ 
		\bm r_{ki}&=\bm r_i-\bm r_k, \quad \mbox{and} \quad 0<\varepsilon_i<1. 
		\end{align*}
	\end{itemize}
	\item[\textbf{Coordination of angular velocities:}] The angular velocity $\omega_i$ of each agent $i$ is governed by the control law:
	\begin{align}
	\omega_i &= -k_{\omega i}\left(\theta_i-\varphi_i\right)+\dot \varphi_i,\label{wi}
	\end{align}
	\noindent where $k_{\omega i}>0$, and $\varphi_i\triangleq\arctan\left(\frac{\F_{iy}}{\F_{ix}}\right)$ is the orientation of a reference vector field $\mathbf F_i$ for agent $i$, defined as:
\end{itemize}
\begin{align}
\label{feedback motion plan}
\mathbf F_i = \prod_{j\in\mathcal N_i} (1-\sigma_{ij}) \mathbf F_{gi} + \sum_{j\in\mathcal N_i} \sigma_{ij} \mathbf F^i_{oj},
\end{align}
where details about the attractive and repulsive vector fields can be found in \cite{Panagou_TAC16}.

Under this protocol we were able to establish collision-free and almost globally convergent motion of the agents towards to their goal locations:

\begin{Multi-Robot}\label{Multi-Robot}
	\textnormal{Consider $N$ agents $i\in\{1,\dots,N\}$ assigned to move to goal locations $\bm r_{gi}$. Then, under the coordination protocol \eqref{velocity protocol}, \eqref{wi}, each agent safely converges to its goal configuration almost globally, except for a set of initial conditions of measure zero.}
\end{Multi-Robot}
\begin{proof}
	The design and analysis of the coordination controller is given in \cite{Panagou_TAC16}.
\end{proof}

In this paper we seek to design a robust coordination protocol so that each agent $i$ can safely accommodate the effects of state and measurement uncertainties $\bm w(t)$, $\bm v_i(t)$, $t\in[0,\infty)$, respectively.
Since we are only concerned about radial convergence of the agents to their respective goal locations, we re-define the radially attractive vector field $\textbf{F}^r_{gi}$ for $\bm r_i\neq \bm r_{gi}$ as:
\begin{subequations}
\begin{align}
\label{Attractive new}
	\F^r_{gix} = \frac{-(x_i-x_{gi})}{(x_i-x_{gi})^2+(y_i-y_{gi})^2},\\ 
	\F^r_{giy} = \frac{-(y_i-y_{gi})}{(x_i-x_{gi})^2+(y_i-y_{gi})^2}.
\end{align}
\end{subequations}
With this globally attractive field, the new reference field is given by \begin{align}
\label{feedback motion plan dynamic}
\mathbf F_i = \prod_{j\in\mathcal N_i} (1-\sigma_{ij}) \mathbf F^r_{gi} + \sum_{j\in\mathcal N_i} \sigma_{ij} \mathbf F^i_{oj}.
\end{align}

\section{Robust Coordination: Design and Analysis}\label{Robust Coordination}
The problem of safe trajectory generation in the presence of disturbances is tackled in two steps. The first step is to modify the nominal controller in order to accommodate for known state disturbances, that in our case can be thought of as wind effects. This task is achieved by feed-forwarding the mean wind speed to each agent's control law as per \eqref{u-p} and \eqref{omega-p}. Once we have a controller that handles this class of known disturbances, the next step is to make it robust \ac{wrt} unknown, zero-mean state disturbances and sensor noises. To accomplish this, an Extended Kalman Filter (EKF) based observer is used to estimate the state vector for the feedback coordination law of each agent. In order to incorporate the estimation error, the crucial safety parameters involved in the nominal control law (i.e., the minimum allowed separation $d_m$ and the set $J_k$ of critical neighbors $k$ to agent $i$) in \eqref{velocity protocol} are modified. Then, with a controller in hand for nominal (or disturbance-free case), feed-forwarded with the known mean wind speed, and by using estimated states as feedback with properly modified safety parameters, we get a robust coordination protocol that steers the agents within a neighborhood of their goal locations, while maintaining safety at all times. Recall that the system is safe if the inter-agent distance $d_{ij}$ between any pair of agents $i$, $j$ is always greater than a minimum allowed separation $d_m$.

\subsection{Control design under bounded disturbances}
We first consider the case where each agent $i$ is subject to known state disturbances, without any measurement uncertainty, i.e., we consider the agent dynamics:
\begin{subequations}
	\label{unicycle-p}
	\begin{align}
	\bm{\dot q}_i&=\bm f(\bm q_i, \bm u^p_i)+\bm \Gamma \bm{\bar w}, \Rightarrow \begin{bmatrix}\dot x_i \\ \dot y_i \\ \dot \theta_i \end{bmatrix}=\begin{bmatrix}u_i^p\cos \theta_i \\u_i^p\sin \theta_i \\ \omega_i^p\end{bmatrix}+ \begin{bmatrix}\bar{w}_{x}\\\bar{w}_y\\0\end{bmatrix},\\
	\bm y_i &= \bm h_i(\bm q_i) = \bm q_i,
	\end{align}
\end{subequations} 
where $\bm{\bar{w}}$ is the known mean of the state disturbance and $\bm u^p_i = \begin{bmatrix} u^p_i & \omega^p_i \end{bmatrix}^T$ is the control input. We propose the following coordination protocol yielding the feedback control law $\bm u^p_i(\bm q,\bm{\bar{w}},d_m)$ for the perturbed dynamics \eqref{unicycle-p} of agent $i$: 

\vspace{2mm}
\noindent \textbf{Coordination of linear velocities:} The linear velocity of each agent $i$ is governed by the control law: 
	\begin{align}
	u_i^p &= \left\{
	\begin{array}{rc}
	-\frac{1}{\mu} \log \Big(\sum \limits_{k\in\mathcal N_i|J_k<0} e^{-\mu u_{i|k}}\Big), & \hbox{$d_m\leq d_{ik}\leq d_{\epsilon}$,}\\
	u_{ic}, & \hbox{$d_{\epsilon}\leq d_{ik}$;} \\
	\end{array}
	\right.,\label{u-p}
	\end{align}
	where: 
	\begin{itemize}
	\item the linear velocity of agent $i$ when free of conflicts is:
	\begin{align}
    u_{ic} &= \left\|u_i\frac{\mathbf F_i}{\|\mathbf F_i\|} - \bm{\bar{w}}\right\| \label{safe velocity}, \\
	u_i &= k_{ui}\tanh\left(\|\bm r_i-\bm r_{gi}\|\right)\label{uin}, 
	\end{align}
	where $k_{ui}>0$ and $\textbf{F}_i$ is given by the nominal vector field \eqref{feedback motion plan dynamic},
		\item for $\bm a = \begin{bmatrix}a_1,\dots ,a_n\end{bmatrix}^T$, the function $g(\bm a) = -\frac{1}{\mu} \log \Big(\sum \limits_{i=1}^N e^{-\mu a_i}\Big)$  is a smooth approximation of the minimum function $\min\{a_1,\dots,a_n\}$ as $\mu \rightarrow \infty $,
		\item $u_{i|k}$ is the safe velocity of agent $i$ \ac{wrt} a neighbor agent $k\in\mathcal N_{i}$, given as:
		\begin{align}
		\label{linear var u}
		u_{i|k}&=u_{ic}\;\frac{d_{ik}-d_m}{d_\epsilon-d_m}+\varepsilon_i \; u_{is|k} \; \frac{d_\epsilon-d_{ik}}{d_\epsilon-d_m},
		\end{align}
		with the terms in \eqref{safe velocity} defined as:
		\begin{align*}
		u_{is|k}&= u^p_k \;\frac{{\bm r_{ki}}^T\bm \eta_k}{{\bm r_{ki}}^T\bm \eta_i},\;\; \bm \eta_i=\left[\begin{matrix}\cos\theta_i\\\sin\theta_i\end{matrix}\right],\;\; J_k={\bm r_{ki}}^T\bm \eta_i, \\ 
		\bm r_{ki}&=\bm r_i-\bm r_k, \quad \mbox{and} \quad 0<\varepsilon_i<1. 
		\end{align*}
	\end{itemize}

\begin{figure}[h]
	\centering
	\includegraphics[width=0.8\columnwidth,clip]{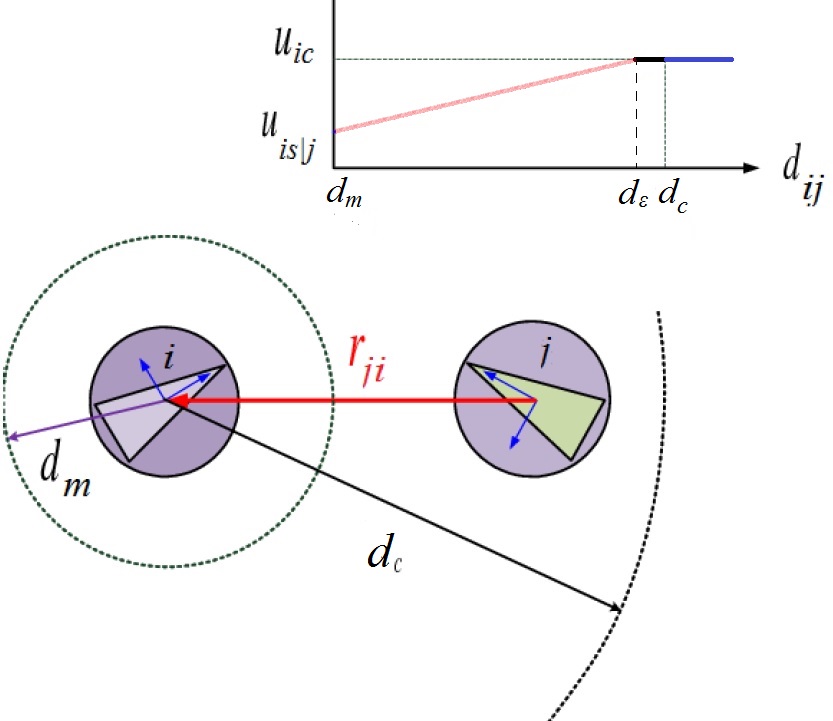}
	\caption{If $J_{j}\triangleq{\bm r_{ji}}^T\bm \eta_i < 0$, i.e., if agent $i$ moves towards agent $j$, then agent $i$ adjusts its linear velocity according to the velocity profile shown here, given analytically by \eqref{linear var u}.}
	\label{fig:velocity adjustment}
\end{figure}

\vspace{2mm}	
\noindent \textbf{Coordination of angular velocities:} The angular velocity of each agent $i$ is governed by the control law:
	\begin{align}
	\omega^p_i &=-k_{\omega i}(\theta_i-\varphi^p_i)+\dot \varphi^p_i, \label{omega-p}
	\end{align}
	where $k_{\omega i}>0$, $\varphi^p_i\triangleq\arctan\left(\frac{\F^p_{niy}}{\F^p_{nix}}\right)$ is the orientation of the normalized vector field $\mathbf F^p_{ni} = \frac{\mathbf F^p_i}{\|\mathbf F^p_i\|}$ for the perturbed system \eqref{unicycle-p} at a point $(x,y)$, with the vector field $\mathbf F^p_i$ for the perturbed system \eqref{unicycle-p} given out of:
	\begin{align}
	\mathbf F^p_i &= u_i\frac{\mathbf F_i}{\|\mathbf F_i\|} - \bm{\bar{w}}, \label{new F}
	\end{align}
	where $\mathbf F_i$ is the nominal vector field given out of \eqref{feedback motion plan dynamic}, and $u_i$ is given out of \eqref{uin}. 
	The time derivative of $\varphi^p_i$ reads 
	\begin{align*}
	\dot \varphi^p_i &=\Big(\left(\begin{matrix}\frac{\partial\F^p_{niy}}{\partial x}\;c\theta^p_i+\frac{\partial\F^p_{niy}}{\partial y}\;s\theta^p_i\end{matrix}\right)\F^p_{nix}\\
	&- \left(\begin{matrix}\frac{\partial\F^p_{nix}}{\partial x}\;c\theta^p_i+\frac{\partial\F^p_{nix}}{\partial y}\;s\theta^p_i\end{matrix}\right)\F^p_{niy}\Big)u_i.    
	\end{align*}
	
It can be readily seen that steady-state solution of $\dot{\theta}_i = \omega^p_i =-k_\omega(\theta_i-\varphi^p_i)+\dot \varphi^p_i$ is $\theta_i = \varphi^p_i$.

\begin{figure}[h]
    \centering
    \includegraphics[width=0.8\columnwidth,clip]{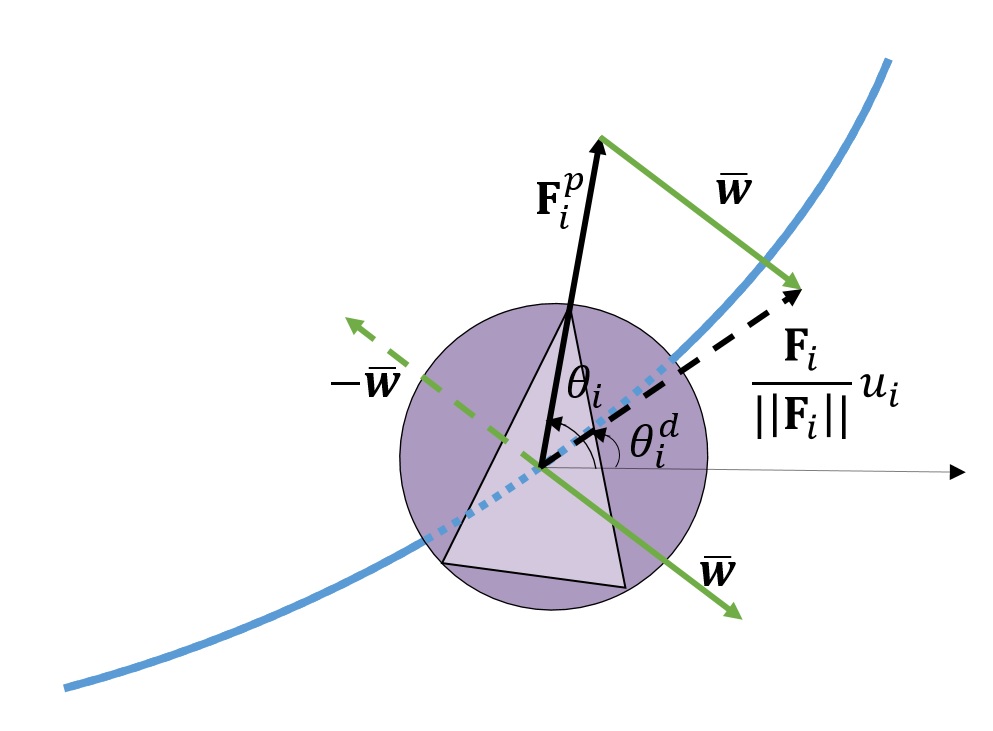}
    \caption{Construction of new vector Field $\textbf{F}^p_i$ in the presence of wind $\bm{\bar{w}}$ to follow the desired vector field $\textbf{F}_i$. Here, blue line is the desired trajectory, green arrow shows the direction of the mean wind speed, dotted black arrow shows direction of nominal vector field $\textbf{F}_i$ while solid orange arrow represents direction of constructed vector field $\textbf{F}^p_i$.}\label{fig:New F graphic}
\end{figure}
        
The following theorem proves that safety for the multi-agent can be guaranteed under the control law given by \eqref{u-p} and \eqref{omega-p}:
\begin{Safety-perturbed} \label{Safety-perturbed}
	If each agent $i\in\{1,\dots,N\}$ subject to the system dynamics \eqref{unicycle-p} follows the control law given by \eqref{u-p} and \eqref{omega-p}, then $\forall i,j\in\{1,\dots,N\}, \; i\neq j$, it holds that:
	\begin{align}
	\|\bm r_i(t)-\bm r_j(t)\| &\geq d_m, \; \forall t \geq 0, \label{safety of perturbed}
	\end{align}
	and each agent converges to its goal location almost globally, i.e.,
	\begin{align}
    \|\bm r_{i\infty}-\bm r_{gi}\| & \triangleq	\lim_{t\to\infty} \|\bm r_i(t)-\bm r_{gi}\| = 0, \label{goal reach}
	\end{align}
except for a set of initial conditions of measure zero.	
\begin{proof}
To prove \eqref{safety of perturbed}, we have that under the control law \eqref{u-p}, agent $i$ adjusts its linear velocity $u_i$ according to the velocity profile shown in Fig. \ref{fig:velocity adjustment}, given analytically out of \eqref{linear var u}, so that the distance $d_{ij}$ \ac{wrt} the worst case neighbor $j$ remains greater than $d_m$. Mathematically, it is required that $$\frac{d}{dt}d_{ij} \big\vert_{d_{ij} = d_m} \geq 0,$$ so that the inter-agent distance does not decrease further once the agents $i$, $j$ are at the minimum allowed separation $d_m$. The derivative of inter-agent distance is given in \eqref{dijdt-long}.
\begin{figure*}
\normalsize
\setcounter{MYtempeqncnt}{\value{equation}}
\setcounter{equation}{16}
\begin{align}
\dot d_{ij} &= \frac{(x_i-x_j)(\dot{x}_i-\dot{x}_j)+(y_i-y_j)(\dot{y}_i-\dot{y}_j)}{\sqrt{(x_i-x_j)^2+(y_i-y_j)^2}}\overset{\eqref{unicycle-p}}{=} \frac{(x_i-x_j)(u^p_ic\theta_i+\bar{w}_x-u^p_jc\theta_j-\bar{w}_x)}{d_{ij}}\nonumber\\
&+\frac{(y_i-y_j)(u^p_is\theta_i+\bar{w}_y-u^p_js\theta_j-\bar{w}_y)}{d_{ij}}=\frac{\Big((x_i-x_j)(u^p_ic\theta_i-u^p_jc\theta_j)+(y_i-y_j)(u^p_is\theta_i-u^p_js\theta_j)\Big)}{d_{ij}}\label{dijdt-long}.
\end{align}
\setcounter{equation}{\value{MYtempeqncnt}}
\hrulefill
\end{figure*}
Hence, we have:
\setcounter{equation}{17}
\begin{align}
		\label{dijdt}
		\frac{d}{dt} \;d_{ij} &=\frac{u^p_i \; {\bm r_{ji}}^T\bm \eta_i - u^p_j \; {\bm r_{ji}}^T\bm \eta_j}{d_{ij}}.
		\end{align} 
		The worst case neighbor for agent $i$ is defined as the agent $j\in \{\mathcal N_{i} \; | \; J_j \triangleq {\bm r_{ji}}^T\bm \eta_i  <0\}$ towards whom the rate of change of inter-agent distance $d_{ij}$ given by \eqref{dijdt}, due to the motion of agent $i$, is maximum. More specifically, the decision making process works as follows: The term $J_j<0$ describes the set of neighbor agents $j$ of agent $i$ towards whom agent $i$ is moving \cite{Panagou_TAC14}. Thus agent $i$ computes safe velocities $u_{i|j}$ \ac{wrt} each neighbor $j\in \{\mathcal N_{i} \; | \; J_j<0\}$, and picks the minimum $u_{i|j}$ of the safe velocities so that the first term in \eqref{dijdt} is as less negative as possible. Now, the value of the safe velocity $u_{i|j}$ \eqref{linear var u} when $d_{ij}=d_m$ is by construction equal to $\varepsilon_i u_{is|k}=\varepsilon_i u_j \frac{{\bm r_{ji}}^T\bm \eta_j}{{\bm r_{ji}}^T\bm \eta_i}$. Plugging this value into \eqref{dijdt} reads: 
		\begin{align}
		\frac{d}{dt}\;d_{ij}=\frac{(\varepsilon_i-1) u_j \; {\bm r_{ji}}^T\bm \eta_j}{d_{ij}}\geq 0.  
		\end{align}
		To see why this condition is true, recall that $\varepsilon_i-1<0$, $u_j\geq 0$, and ${\bm r_{ji}}^T\bm \eta_j\leq 0$: this is because agent $j$ is either following a vector field $\mathbf F_j$ that points away from agent $i$, or happens to move away from agent $i$ in the first place. This implies that the inter-agent distance $d_{ij}$ can not become less than $d_m$ \cite{Panagou_TAC16}. To verify this argument, it is sufficient to show that in the presence of known disturbances $\bm{\bar{w}}$, the motion of any agent $i,j\in\{1,2,\dots,N\}$ under the control law \eqref{u-p} and \eqref{omega-p} is along their nominal vector fields given by \eqref{feedback motion plan dynamic}, as follows: 
		
		Let $\theta^d_i$ be the nominal direction that agent $i$ is supposed to follow under the nominal vector field $\mathbf F_i$, i.e., $\theta^d_i \triangleq \arctan \left( \frac{\F_{ix}}{\F_{iy}}\right)$. From the steady-state solution of $\dot{\theta_i} = \omega_i$ under the control law \eqref{omega-p}, we have $\theta_i = \varphi^p_i$, where $\theta_i$ is the orientation of the agent $i$. 
		
		Let $\angle{(\cdot)}$ be signed angle, defined to be positive in the clockwise direction and negative in the counter-clockwise direction. Now, from \eqref{unicycle-p} we have: $\angle{(\dot{\bm r}_i-\bm{\bar{w}})} = \theta_i$. Also, by construction of the new vector field \eqref{new F}, $\angle \left(u_i\frac{\textbf{F}_i}{\|\textbf{F}_i\|}-\bm{\bar{w}}\right) = \varphi^p_i \overset{\eqref{omega-p}}{=} \theta_i$ out of the steady-state solution of \eqref{omega-p}, see also Figure \ref{fig:New F graphic}. Thus we have that $\angle{(\dot{\bm r}_i-\bm{\bar{w}})} = \angle \left(u_i\frac{\textbf{F}_i}{\|\textbf{F}_i\|}-\bm{\bar{w}}\right)$, which makes $\angle {\dot{\bm r_i}} = \angle{\textbf{F}_i} = \theta^d_i$. Hence, the motion of agent $i$ is along the desired nominal vector field $\mathbf F_i$. Analysis of convergence of the agents to their goal locations is given in \cite{Panagou_TAC16}. Note that the analysis given in \cite{Panagou_TAC16} is carried out for dipole type attractive vector field, which will still hold for the radially attractive vector field. Since, except for a set of initial conditions of measure zero, the nominal vector field drives the agents to their goal location (Theorem \ref{Multi-Robot}), it follows that the modified coordination protocol also drives the agents to their goal locations almost globally. 
	\end{proof}

	It is to be noted that while the direction of motion of agent $i$ is along the nominal vector field $\textbf{F}_i$, its orientation (heading angle) $\theta_i$ remains along the modified vector field $\textbf{F}_i^p$. 
\end{Safety-perturbed}

\subsection{Extension of the controller for stochastic disturbances}

In order to estimate the states of the system in the presence of stochastic disturbances, we design a system observer based on the Extended Kalman Filter:
\begin{subequations}
	\label{Estimation}
	\begin{align}
	\dot{\bm {\hat{q}}}_i &= \bm f(\bm{\hat{q}}_i,\bm u_i)+\bm \Gamma \bm{\bar{w}} + \bm{K}_i(\bm y_i-\hat{\bm y}_i),\label{EKF estimate}\\
	\dot{\bm P}_i &= \bm A_i \bm P_i+ \bm P_i \bm A_i^T -\bm{K}_i\bm H_i\bm P_i + \bm \Gamma\bm P_w\bm \Gamma^T, \\
	\hat{\bm y}_i &= \bm h(\bm {\hat{q}}_i) = \bm {\hat{q}}_i, \label{measurement model} \\
	\bm{K}_i &= -\bm P_i\bm H_i^T\bm P_{v_i}^{-1}, \\
	\bm u_i&= \bm u_i(\bm{\hat{q}}, \bm{\bar{w}}, d_m',\epsilon_J), \label{control hat}
	\end{align}
\end{subequations}
where $\bm A_i = \left.\frac{\partial \bm f}{\partial \bm q_i } \right \vert _{\hat{\bm{q}}_i}$ is the state matrix of the linearized dynamics evaluated at $\hat{\bm q}_i$, 
$\bm H_i = \left . \frac{\partial \bm h}{\partial \bm{q_i} } \right \vert _{\hat{\bm{q}}_i}$ is the linearized output matrix evaluated at $\hat{\bm q}_i$, $\bm{P}_i$ is the covariance matrix of the estimation error $\tilde{\bm q_i}=\bm q_i - \hat{\bm q}_i$, and $\bm{K}_i$ is the Kalman gain. Note that the control law \eqref{control hat} has the same form as \eqref{u-p} but it uses the estimated states $\hat{\bm q}$ for feedback, the mean wind speed $\bm{\bar{w}}$ for feed-forward command, and involves $d_m'$ as the new safety parameter. The last parameter $\epsilon_J$ is introduced to re-define the worst-case neighbor, as per \eqref{worst case new}.

\textbf{Stability of EKF-based estimator:} In \cite{reif2000stochastic}, authors have shown that the EKF is stable if there exist positive constants $\bar c, \underline p, \underline q, \delta, k$ such that:
\begin{itemize}
    \item $\|\bm H(t)\| \leq \bar c$
    \item System is uniformly observable
    \item $\underline p \leq \bm P_{v} \leq \delta\bm I$
    \item $\underline q \leq \bm P_{w} \leq \delta \bm I$
    \item $\|H.O.T\| \leq k\|\tilde{\bm q}_i\|^2$
\end{itemize}
where $\tilde{\bm q}_i = \bm q_i - \hat{\bm q}_i$ and $H.O.T$ stands for higher-order terms of the linearization . Furthermore, if these conditions hold, then the estimation error is exponentially mean-square bounded. Using this result, we now show that EKF based estimator will remain bounded for system \eqref{unicycle} whose control law is given by \eqref{control hat}. 
In our system, $\bm H$ is just an identity mapping, which allows us to choose $\bar c = 1$. Since $\bm H$ is identity, we have that the system is always observable. By the assumption on the type of disturbance we consider in \eqref{unicycle}, the covariance matrices $\bm P_{v_i}$ and $\bm P_w$ are non-zero constant matrices and hence are bounded. We assume that the initial estimation error is bounded. Now, we show that our closed-loop system also satisfies the condition on boundedness of $H.O.T$. The function $\bm f$ of the system dynamics given by \eqref{unicycle} is continuously differentiable in $\bm q_i$ and $\bm u_i$. From \eqref{u-p} and \eqref{omega-p} we have that the control law \eqref{control hat} is a bounded, continuously differentiable function of $\bm q$, since the vector field $\textbf{F}^p_i$, its partial derivatives \ac{wrt} $x_i$ and $y_i$, and the linear control input $u^p_i$, are continuously differentiable and bounded. Hence for all agents the closed loop function $\bm f(\bm q_i)$ is a continuously differentiable, bounded function of the states of the agent $i$. Hence all its derivatives $\frac{d^k\bm f}{d\bm q^k_i}$ are bounded in $\mathbb R^n$. Using the expression for Lagrange Remainder for Taylor series expansion \cite{abramowitz1964handbook}, since we are using first order expansion, we have that the norm of the remainder of Taylor series expansion or the higher order terms $\|H.O.T\|\leq \frac{1}{2}\|\frac{d^2 \bm f}{d \bm q^2_i}\| \|\tilde{\bm q}_i\|^2$ and $\|\frac{d^2 \bm f}{d \bm q^2_i}\|\leq L$. Hence $\|H.O.T\| \leq k\|\tilde{\bm q}_i\|^2$ with $k = L/2$.
 
Thus, the EKF-based estimator \eqref{Estimation} is stable, which implies that the estimation error $\tilde{\bm q}$ will remain bounded for sufficiently small disturbances, i.e., for small covariance matrices $\bm P_w$  and $\bm P_{v_i}$. We can therefore bound the norms of the estimation errors in individual states and position of agent $i$ by small, positive numbers. Define the maximum of the errors in the estimation of the states for any agent $i$ as $\epsilon_x \triangleq 3\max\limits_i\sqrt{\bm P_{i11}} $, $ \epsilon_y \triangleq 3\max\limits_i\sqrt{\bm P_{i22}} $ and $ \epsilon_\theta \triangleq 3\max\limits_i\sqrt{\bm P_{i33}}$ where $\bm P_{ill}, l\in\{1,\dots,n\}$, are the diagonal terms of error covariance matrix $\bm P_i$. Also, we define the maximum error in the estimation of the position as $\epsilon_d \triangleq \sqrt{\epsilon_x^2+\epsilon_y^2}$. Using the 3-$\sigma$ bound for the Gaussian noise, we have that $\|x_i-\hat{x}_i\| \leq \epsilon_x $, $ \|y_i-\hat{y}_i\| \leq \epsilon_y $, $ \|\theta_i-\hat{\theta_i}\| \leq \epsilon_{\theta}$ and $\|\bm r_i-\hat{\bm r}_i\| \leq \epsilon_d$. Referring to Definition 4.1 of \cite{reif2000stochastic}, we can choose a function $\Lambda(\bm q) = -1$, so that $\|\Lambda\| = 1$. Then, using Theorem 7 in \cite{baras1988dynamic}, we have that $\|\bm P_i(t)\| \leq \|\bm P_i(0)\| + 1$. We choose $\bm P_i(0) = \bm P_{v_i}$, so that $\max \|\bm P(t)\| = \max\limits_i\|\bm P_{v_i}\| +1$. With this bound on the covariance matrix, we can express $\epsilon_x = \epsilon_y = \epsilon_{\theta} = \sqrt{\max \limits_i \|\bm P_{v_i}\|+1}$ and $\epsilon_d = \sqrt{2 \left(\max \limits_i \|\bm P_{v_i}\|+1\right)}$. 

\subsection{Safety Analysis}
The linear velocity coordination law \eqref{u-p} depends upon the safety parameter $d_m$. As only estimates of the inter-agent distances are available to each agent, one can set the minimum allowed estimated distance $\hat d_{ij_{min}}$ equal to $d_m'$ (given by \eqref{new param 1}), in order to ensure that actual inter-agent distance $d_{ij}$ remains greater than the minimum allowed distance $d_m$. 

We re-define the worst case neighbor (defined earlier) in terms of the estimated states as: 
\begin{align}
	\label{worst case new}
	j &\in \{\mathcal N_{i}\;|\; {\bm \hat{\bm r}_{ji}}^T\bm \hat{\bm \eta}_i \leq-\epsilon_J\}.
	\end{align}
To proceed with the analysis, we first prove the following Lemma: 
\begin{Safe worst case} \label{new worst}
	If $\bm{\hat{\bm r}}^T_{ji}\bm \hat{\bm \eta}_i \leq-\epsilon_J$, where $\epsilon_J = \frac{2\epsilon_d+s(\epsilon_{\theta})(d_m+2\epsilon_d)}{c(\epsilon_{\theta})}$, then with probability 0.997
	\begin{align*}
	{\bm r^T_{ji}}\bm \eta_i &\leq 0.
	\end{align*}
\begin{proof}
Since we are concerned about safety of the system, let's assume that inter-agent distance $d_{ij} = d_m$. We use the fact that $\|x_i-\hat{x}_i\| \leq \epsilon_x $, $ \|y_i-\hat{y}_i\| \leq \epsilon_y $, $ \|\theta_i-\hat{\theta_i}\| \leq \epsilon_{\theta}$ and $\|\bm r_i-\hat{\bm r}_i\| \leq \epsilon_d$. Therefore we have that 
\begin{align*}
d_m-2\epsilon_d\leq \|\hat{\bm r}_i-\hat{\bm r}_j\|\leq d_m+2\epsilon_d.
\end{align*}
Now, 		
\begin{align*}
{\bm r_{ji}}^T\bm \eta_i &= {\bm r_{ji}}^T\left[\begin{matrix}\cos\theta_i\\\sin\theta_i\end{matrix}\right]= (x_i-x_j)c\theta_i + (y_i-y_j)s\theta_i,
\end{align*}
which further reads as in \eqref{derivs}.
\begin{figure*}
\normalsize
\setcounter{MYtempeqncnt}{\value{equation}}
\setcounter{equation}{21}
\begin{align}
{\bm r_{ji}}^T\bm \eta_j &= \bigl((x_i-\hat{x}_i)-(x_j-\hat{x}_j)+(\hat{x}_i-\hat{x}_j)\bigr)c\theta_i + \bigl((y_i-\hat{y}_i)-(y_j-\hat{y}_j)+(\hat{y}_i-\hat{y}_j)\bigr)s\theta_i \nonumber\\
&\leq (\hat{x}_i-\hat{x}_j)c\theta_i + (\hat{y}_i-\hat{y}_j)s\theta_i +2\epsilon_xc\theta_i + 2\epsilon_ys\theta_i \leq (\hat{x}_i-\hat{x}_j)c(\hat{\theta_i} +\epsilon_{\theta_i}) + (\hat{y}_i-\hat{y}_j)s(\hat{\theta_i}+\epsilon_{\theta_i}) +2\epsilon_d \nonumber \\
& \leq (\hat{x}_i-\hat{x}_j)\bigl(c\hat{\theta_i}c(\epsilon_{\theta})-s\hat{\theta_i}s(\epsilon_{\theta})\bigr) +(\hat{y}_i-\hat{y}_j)\bigl(s\hat{\theta_i}c(\epsilon_{\theta})+c\hat{\theta_i}s(\epsilon_{\theta})\bigr) + 2\epsilon_d \nonumber \\
&= \left((\hat{x}_i-\hat{x}_j)c\hat{\theta_i}+ (\hat{y}_i-\hat{y}_j)s\hat{\theta_i}\right)c(\epsilon_{\theta})-s(\epsilon_{\theta})\left((\hat{x}_i-\hat{x}_j)s\hat{\theta_i}- (\hat{y}_i-\hat{y}_j)c\hat{\theta}_i\right) +2\epsilon_d \nonumber \\
& \leq \left((\hat{x}_i-\hat{x}_j)c\hat{\theta}_i+ (\hat{y}_i-\hat{y}_j)s\hat{\theta}_i\right)c(\epsilon_{\theta})+s(\epsilon_{\theta})(d_m+2\epsilon_d)+2\epsilon_d\label{derivs}
\end{align}
\setcounter{equation}{\value{MYtempeqncnt}}
\hrulefill
\end{figure*}
\setcounter{equation}{22}
Let $\epsilon_J = \frac{2\epsilon_d+s(\epsilon_{\theta})(d_m+2\epsilon_d)}{c(\epsilon_{\theta})}$. Since $\epsilon_{\theta}$ and $\epsilon_d$ are small positive numbers, we have $\cos(\epsilon_{\theta})>0$ and $\epsilon_J>0$. Hence, if $((\hat{x}_i-\hat{x}_j)c\hat{\theta}_i+ (\hat{y}_i-\hat{y}_j)s\hat{\theta}_i)\leq-\epsilon_J$, then we have that $\bm r_{ji}^T\bm \eta_i \leq 0$. Since the bounds on the estimation error are probabilistic, we only guarantee that this result holds with probability 0.997\footnote{Note that this can be increased to a value arbitrarily close to 1 by using k-$\sigma$ rule to bound the estimation error, where k is arbitrarily large.}.     
\end{proof}
\end{Safe worst case}

\vspace{2mm}
Now we use this result to show the safety of the system. In order to maintain the safety of the agents in the presence of disturbances, the safety parameter $d_m'$ in the control law \eqref{control hat} is given by:
\begin{align}
\label{new param 1}
d_{m}' &= d_{m}+2\epsilon_{d}.
\end{align}

\begin{Safety-stoch}
If each agent $i$ under the system dynamics \eqref{unicycle} follows the control law given by \eqref{control hat}, and if $d_m'$ is given by \eqref{new param 1}, then with probability 0.997, $\forall t \geq 0, \forall \; i, j\in\{1,\dots,N\}, \; j\neq i$:
\begin{align}
	\label{safety of actual}
	\|\bm r_i(t)-\bm r_j(t)\| \geq d_m.
	\end{align}
	\begin{proof}
	    Using triangle inequality,  
		\begin{align*}
		\|\hat{\bm r}_i-\hat{\bm r}_j\| &\leq \|\hat{\bm r}_i-\bm r_i\| + \|\hat{\bm r}_j-\bm r_j\| + \|\bm r_i-\bm r_j\| \Rightarrow \\
		\|\hat{\bm r}_i-\hat{\bm r}_j\| &\leq 2\epsilon_d + \|\bm r_i-\bm r_j\| \Rightarrow d_m' \leq 2\epsilon_d+\|\bm r_i-\bm r_j\|. 
		\end{align*}
		Choosing $d_m'$ as per \eqref{new param 1}, we get $\|\bm r_i-\bm r_j\| \geq d_m$.
		To complete the proof, it is sufficient to show that when $\hat{d}_{ij} = d_m'$, the time derivative of inter-agent distance is positive, i.e. $\dot{d}_{ij}>0$. 
		In the presence of uncertainties, to guarantee that the system is safe, we modify the definition of the worst case neighbor as $j \in \{\mathcal N_{i} \;|\; \hat{J}_j \triangleq \bm{\hat{\bm r}}^T_{ji}\bm \hat{\bm \eta}_j <-\epsilon_J\}$ so that, from Lemma \ref{new worst} we have that ${\bm r^T_{ji}}\bm \eta_j \leq 0$, which implies $\dot{d}_{ij} \geq 0$ at $\hat{d}_{ij} \triangleq \|\hat{\bm r}_i-\hat{\bm r}_j\| = d_m'$. 
		Therefore, we have that: (i) $\hat{d}_{ij} = d_m' \implies d_{ij} \geq d_m$, and (ii) $\frac{d}{dt}d_{ij} \big\vert_{\hat{d}_{ij} = d_m'} \geq 0$, i.e. the inter-agent distance does not decrease beyond this point. Hence the multi-agent system always remains safe. 
	\end{proof}
\end{Safety-stoch}

\subsection{Analysis of convergence to goal location}
\begin{Convergence}	
If agent $i$ under the system dynamics \eqref{unicycle} follows the control law given by \eqref{control hat}, then with probability 0.997:
\begin{align}
\label{conv of actual}
\lim_{t\to\infty} \|\bm r_i(t)-\bm r_{gi}\|&= \|\bm r_{i\infty}-\bm r_{gi}\| \leq \epsilon_f,
\end{align}
where $\epsilon_f = \epsilon_d + \epsilon$, where $\epsilon$ is an arbitrary small positive constant. 

\begin{proof}Using triangle inequality, 
	\begin{align*}
	\|\bm r_{i\infty}-\bm r_{gi}\| &\leq \|\bm r_{i\infty}-\hat{\bm r}_{i\infty}\| + \|\hat{\bm r}_{i\infty}-\bm r_{gi}\|.
	\end{align*} 
	System \eqref{EKF estimate} can be seen as a perturbed form of \eqref{unicycle-p} with $\bm y_i -\hat{\bm y}_i$ being the constantly acting, bounded perturbation. From Theorem \ref{Safety-perturbed}, we have that $\bm r_{gi}$ is an (almost globally) asymptotically stable equilibrium of \eqref{unicycle-p}. Furthermore, (assuming no interactions among agents in the vicinity of the goal locations) one has that $\bm r_{gi}$ is a stable equilibrium of \eqref{EKF estimate}. To verify this argument, use the candidate Lyapunov function $V= \|\hat{\bm r}_i - \bm r_{gi}\|^2$. Define $\bm r_e \triangleq \hat{\bm r}_i - \bm r_{gi}$. 

   Now, taking the time derivative of candidate Lyapunov function along the system \eqref{EKF estimate}, we get
    \begin{align*}
	    \dot{V} &= \bm r_e^T\dot{\hat{\bm r}}_i = \bm r_e^T\left[\begin{matrix}u^p_ic\varphi^p_i + \bar{w}_x  \\ u^p_is\varphi^p_i +\bar{w}_y\end{matrix}\right] 
	     + \bm r_e^T\bm\Gamma^T\bm{K}_i(\bm y_i-\hat{\bm y}_i).
	    \end{align*}
	    Note that $\left[\begin{matrix}u^p_ic\varphi^p_i + \bar{w}_x \\u^p_is\varphi^p_i +\bar{w}_y\end{matrix}\right]$ is the $\dot{\bm r}^p_i$ of agent $i$ in the absence of unknown disturbance with magnitude $|\dot{\bm r}^p_i| = u_i$ and is along the attractive vector field $\textbf{F}^r_{gi}$ which points in the direction $-(\hat{\bm r}_i-\bm r_{gi})$. Hence, we have that  $\left[\begin{matrix}u^p_ic\varphi^p_i + \bar{w}_x \\u^p_is\varphi^p_i +\bar{w}_y\end{matrix}\right] = -u_i\frac{\bm r_e}{\|\bm r_e\|}$. Finally, since the EKF based estimator is stable, the perturbation term $\bm{K}_i(\bm y_i-\hat{\bm y}_i)$ can be bounded by $\|\bm K_i\|\epsilon_d \leq \delta$. Hence, the time derivative $\dot V$ reads 
	 \begin{align*}
	    \dot V &= -\bm r_e^Tu_i\frac{\bm r_e}{\|\bm r_e\|} + \bm r_e^T\bm\Gamma^T\bm{K}_i(\bm y_i-\hat{\bm y}_i) \\
	    &\leq -k_{ui}\|\bm r_e\|\tanh(\|\bm r_e\|) + \delta\|\bm r_e\|.
	\end{align*} 
	   Hence, we have $\dot{V}\leq 0$ for $\|\bm r_e\| = \|\hat{\bm r}_i - \bm r_{gi}\| \geq \tanh^{-1}\frac{\delta}{k_{ui}}$ where $k_{ui}$ is chosen to be greater than $\delta$. Using the stability of perturbed systems under the effect of constantly acting (non-vanishing) perturbations \cite{khalil1996noninear}, we have that $\bm r_{gi}$ is a stable equilibrium of \eqref{EKF estimate}, which ensures that $\|\hat{\bm r}_{i\infty}-\bm r_{gi}\| \leq \epsilon$ for some small positive number $\epsilon>0$. Therefore,  
\begin{align*}
	\|\bm r_{i\infty}-\bm r_{gi}\| \leq \|\bm r_{i\infty}-\hat{\bm r}_{i\infty}\| + \epsilon \Rightarrow \|\bm r_{i\infty}-\bm r_{gi}\| \leq \epsilon_d +\epsilon.
	\end{align*}
	Choosing $\epsilon_f = \epsilon_d +\epsilon$ completes the proof.
\end{proof}
\end{Convergence}

It is important to note that in the presence of stochastic state-disturbance it cannot guaranteed that an agent can achieve any given goal orientation. In fact, the steady-state orientation of any agent would be opposite to the wind direction with small deviations because of uncertain state disturbance. This is true because at goal location, $\textbf{F}_i = \bm 0$ and hence, $\textbf{F}^p_i$ is along the opposite direction of wind.  

\section{Simulations}\label{Simulations}
We consider two scenarios involving $\N=20$ agents which are assigned to move towards goal locations while avoiding collisions. The goal locations are selected sufficiently far apart so that the agents' communication regions do not overlap when agents lie on their goal locations. The covariance matrix of the state disturbance is taken as $\bm P_w = \mathrm{diag}(0.01, 0.01)$ and measurement noise as $\bm P_{vi} = \mathrm{diag}(0.01, 0.01, 0.01)$. In the first case, we assume the mean wind speed to be constant with time, with $\bm{\bar{w}} = \begin{bmatrix}-0.2& 0.7\end{bmatrix}^T (m/sec)$. With these uncertainties, we have $\epsilon_d = 1.42\;m$ and $\epsilon_f = 1.52 \;m$. The radii of the agents are $\varrho=0.4\;m$, the minimum separation is set equal to $d_m=2 \varrho=0.8\;m$, and the communication radius is set equal to $R_c= 2d_m' = 7.28\;m$. 
\begin{figure}[h]
	\centering
	\includegraphics[width=1\columnwidth,clip]{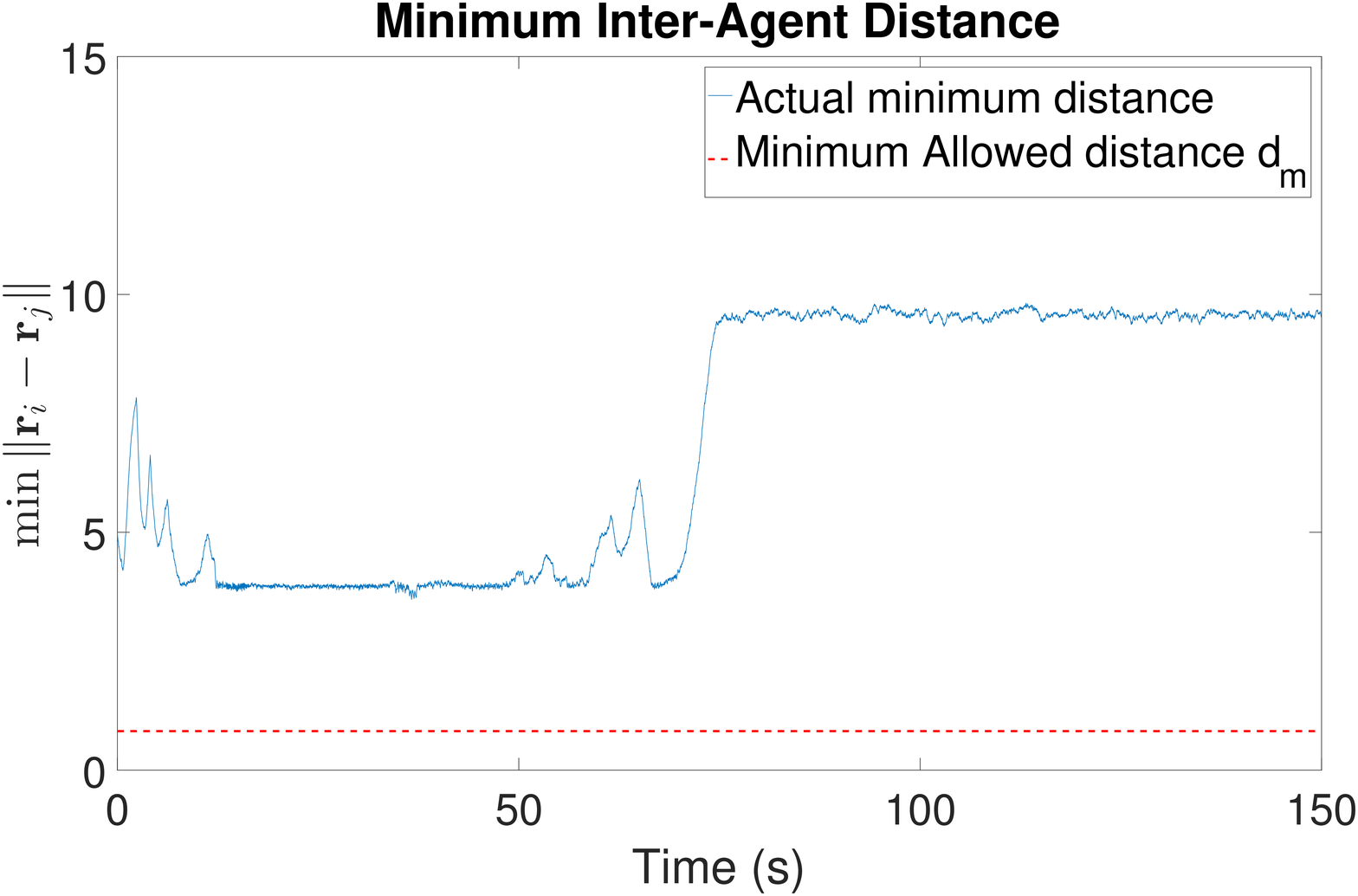}
	\caption{The smallest pairwise distance at each time instant for constant $\bar{\bm w}$.}
	\label{fig:min dist cons w}
\end{figure}

\begin{figure}[h]
	\centering
	\includegraphics[width=1\columnwidth,clip]{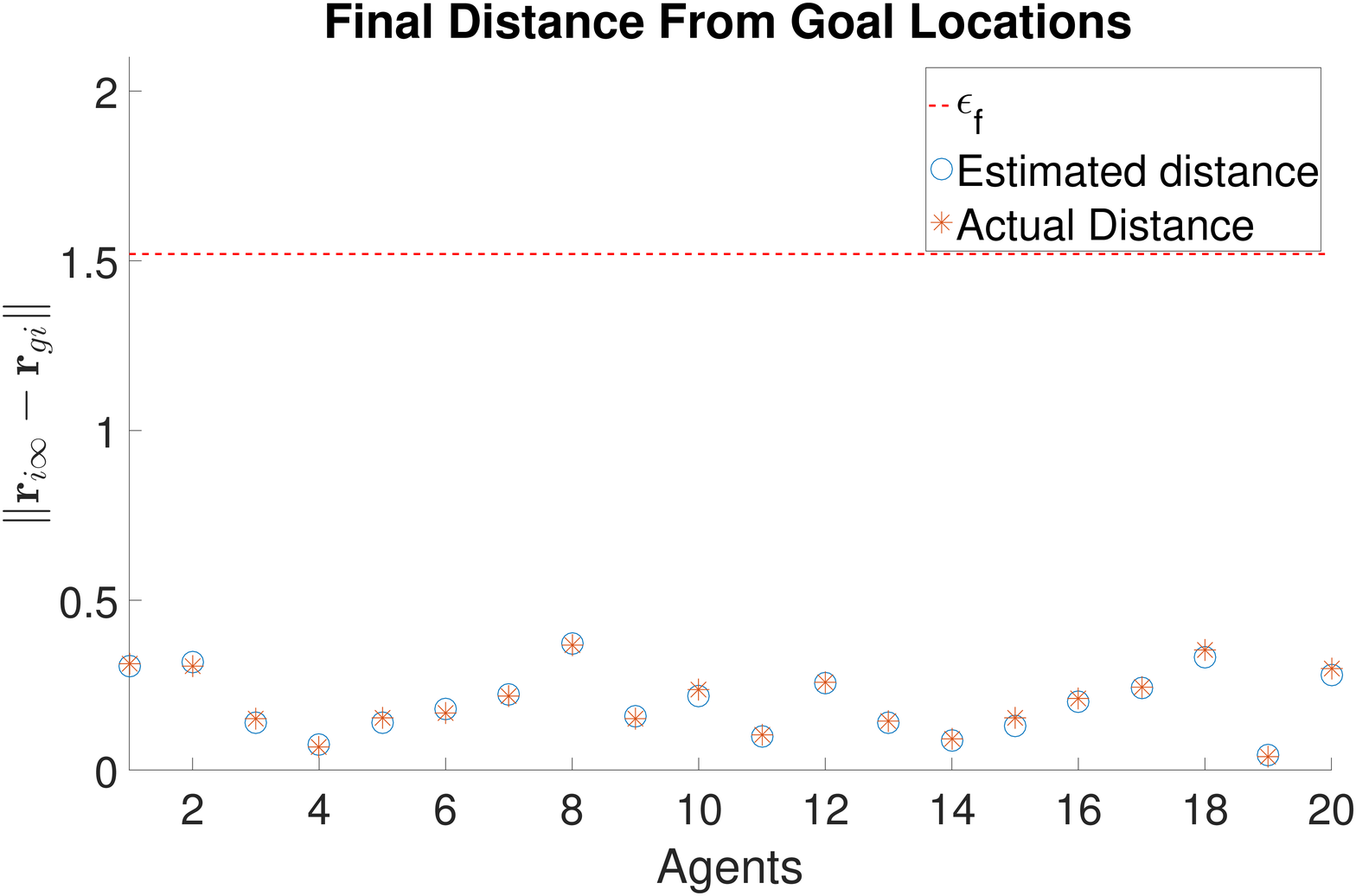}
	\caption{Final distance from the goal location for constant $\bar{\bm w}$.} 
	\label{fig:fin dist cons w}
\end{figure}

\begin{figure}[h]
	\centering
	\includegraphics[width=1\columnwidth,clip]{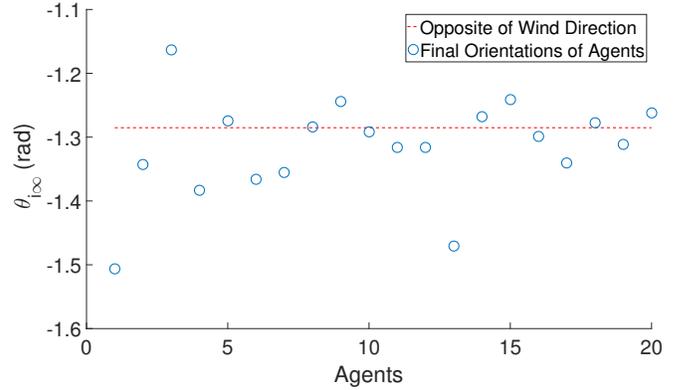}
	\caption{Final orientations $\theta_{i\infty}$ and opposite direction of the wind. }
	\label{fig:fin orient}
\end{figure}

\begin{figure}[h]
	\centering
	\includegraphics[width=1\columnwidth,clip]{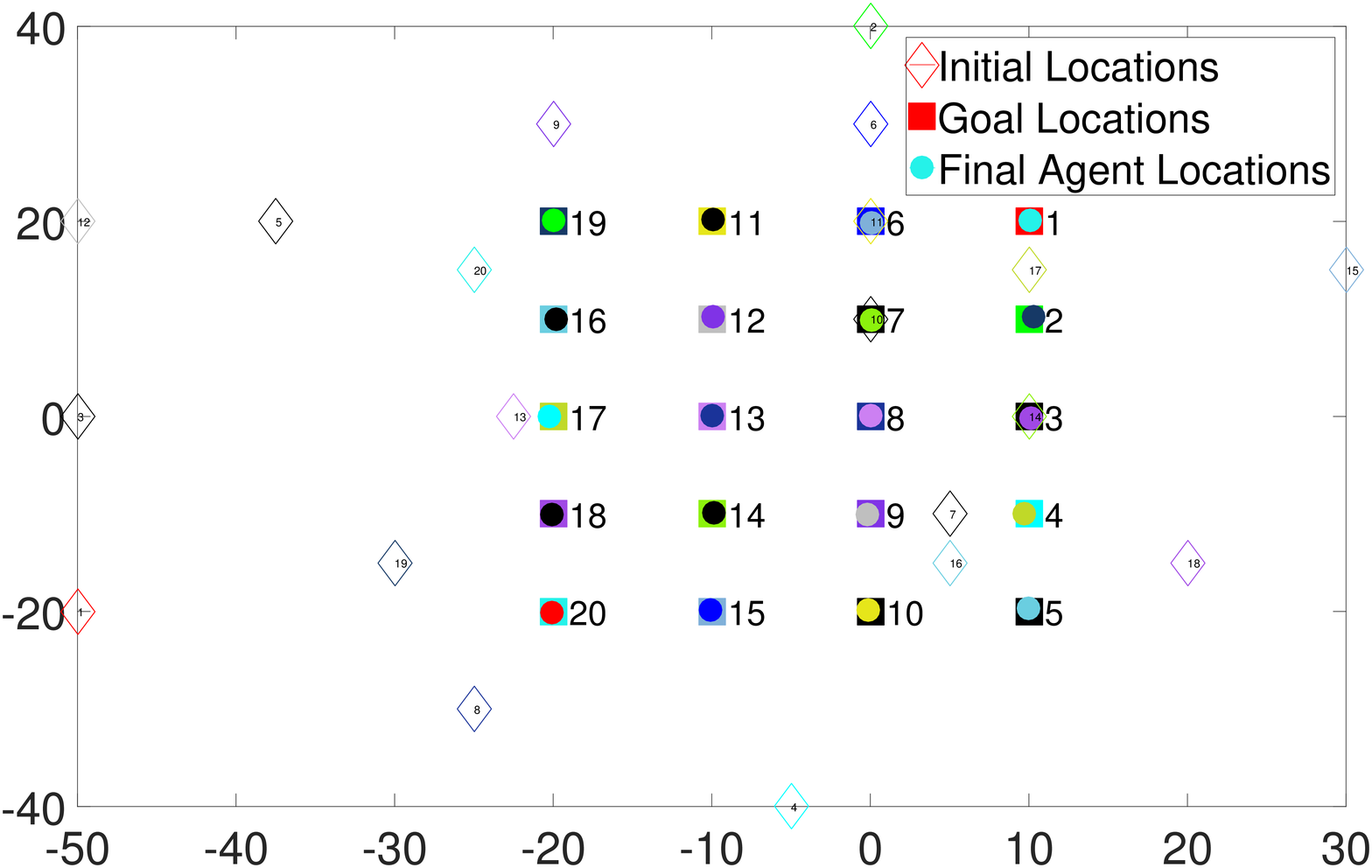}
	\caption{Initial, goal locations and the positions reached by the agents after 150 seconds (15,000 iterations).}
	\label{fig:init final const w}
\end{figure}

Figures \ref{fig:min dist cons w} and \ref{fig:fin dist cons w} respectively show the minimum distance between any pair of agents and the final distances of agents from their respective goal locations. It can be seen that the agents reach the $\epsilon_f$ ball around their respective goals, while maintaining the $d_m$ distance between each other. Figure \ref{fig:fin orient} shows that agents align themselves opposite to the wind direction in the equilibrium. Figure \ref{fig:init final const w} shows the initial, goal and the final positions reached by agents after 15,000 iterations with time-step $0.01\;sec$. 

\begin{figure}[h]
	\centering
	\includegraphics[width=1\columnwidth,clip]{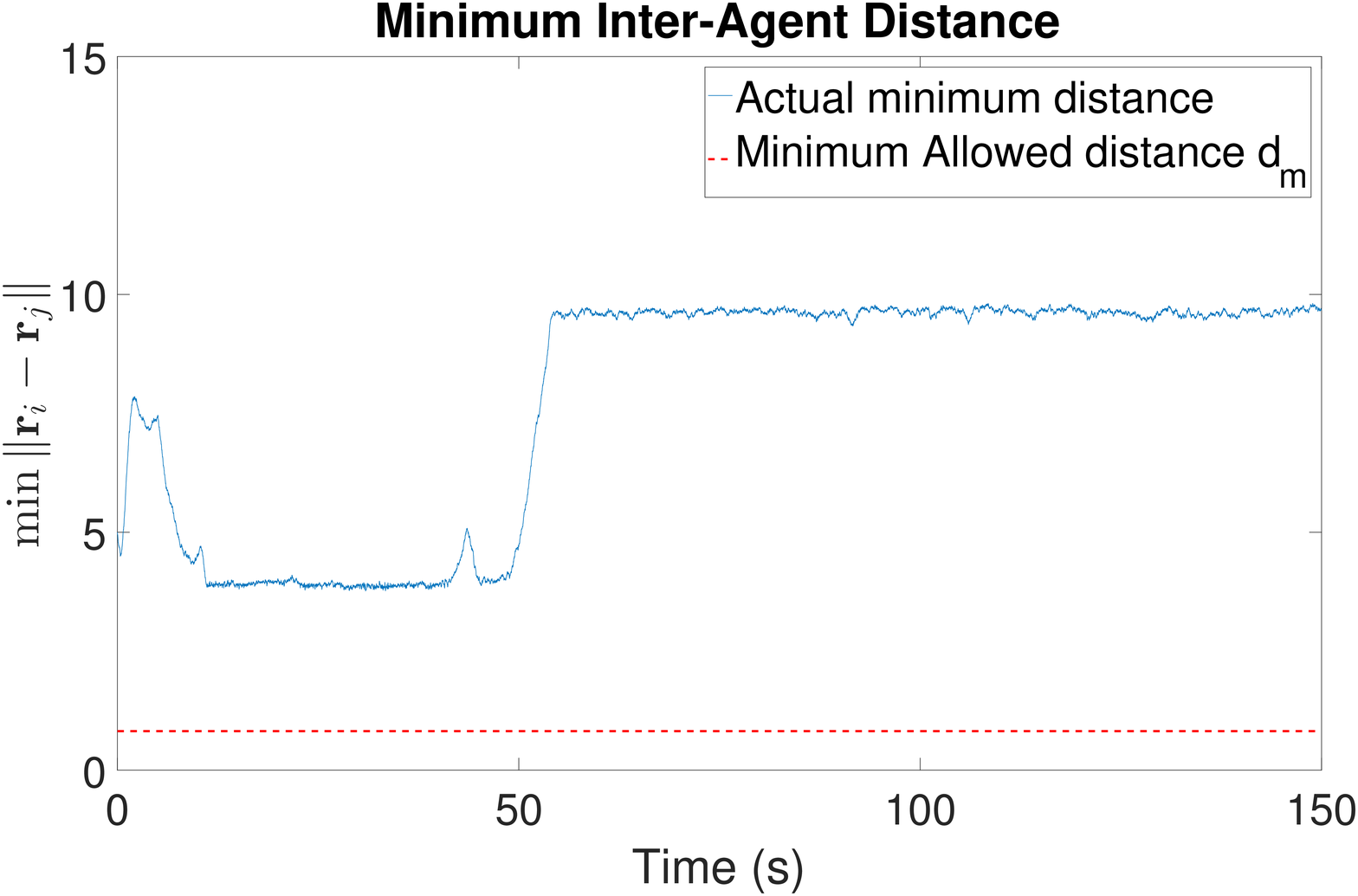}
	\caption{The smallest pairwise distance at each time instant for time varying $\bar{\bm w}$.}
	\label{fig:min dist}
\end{figure}

\begin{figure}[h]
	\centering
	\includegraphics[width=1\columnwidth,clip]{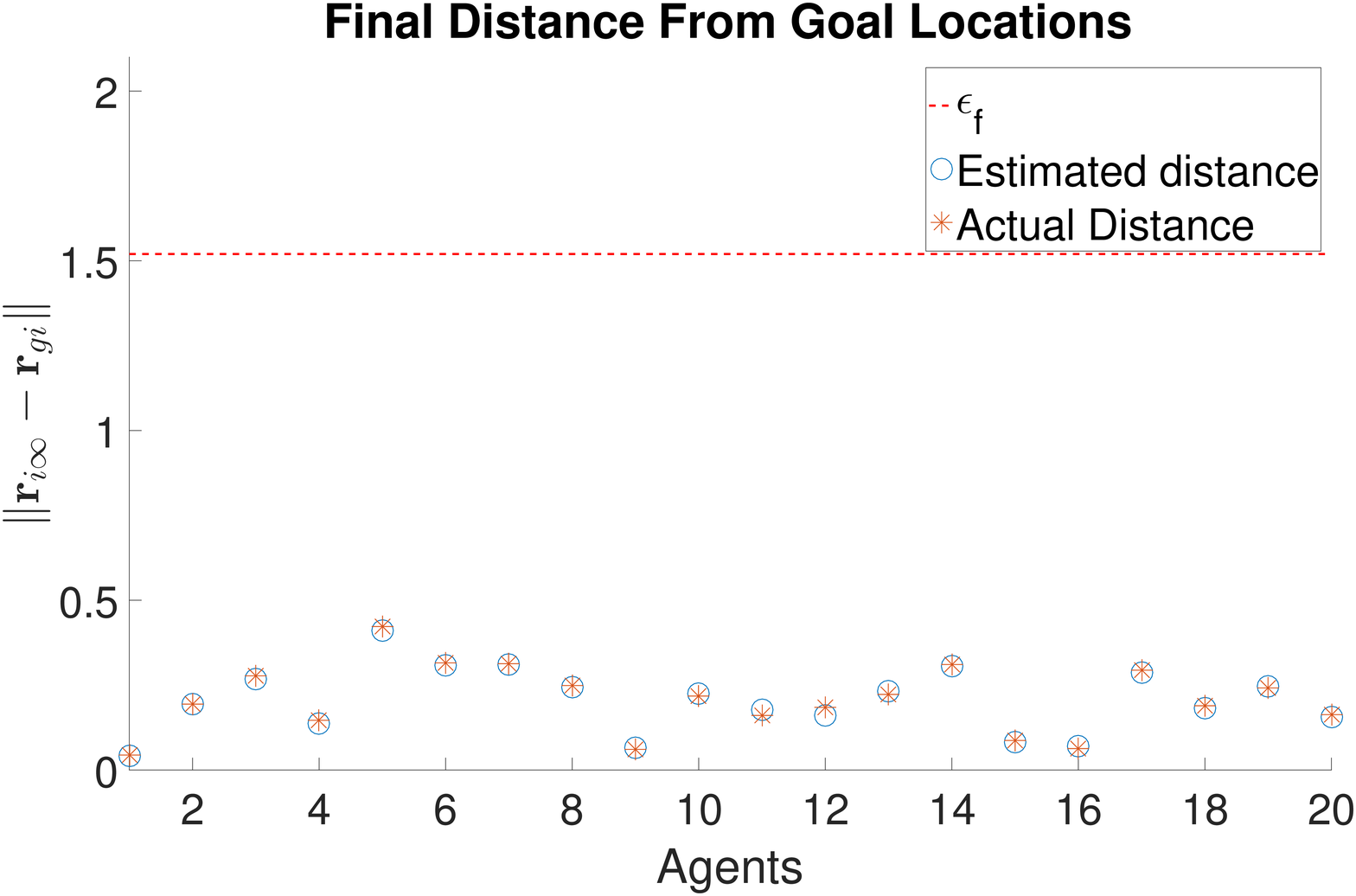}
	\caption{Final distance from the goal location for time varying $\bar{\bm w}$.} 
	\label{fig:fin dist}
\end{figure}

In the second case, we assume the mean wind speed to be time varying with $\bar{w}_x(t),\bar{w}_y(t)\in [-1, 1]\;(m/sec)$. Similar to the first case, plots are given to show the performance of the coordination protocol in presence of time-varying wind disturbance. Figure \ref{fig:min dist} and \ref{fig:fin dist} depict the minimum pairwise distances between the agents and their final distance from their goal locations, respectively.

\section{Conclusions and Future Work}\label{Conclusions}

We presented a safe semi-cooperative multi-agent coordination protocol under state and measurement uncertainty. The nominal case of our earlier work is redesigned by feed-forward control, vector-field-based feedback control, and nonlinear estimation techniques, so that safety and convergence of the agents up to some bound around the desired destination is guaranteed. In the future, we would like to study the case when the mean state disturbance has a spatial distribution. 
Ongoing work focuses on treating the case of input- and state- constrained agents under uncertainties, with application to fixed-wing aircraft operating in obstacle environments. 

\bibliographystyle{IEEEtran}
\bibliography{myreferences}

\end{document}